\documentclass[%
 aip,
 amsmath,amssymb,
 reprint,%
]{revtex4-2}

\usepackage{graphicx}% Include figure files
\usepackage{dcolumn}% Align table columns on decimal point
\usepackage{bm}% bold math
\usepackage[utf8]{inputenc}
\usepackage{float}
\usepackage{amssymb}
\usepackage[T1]{fontenc}
\usepackage{color,soul}
\usepackage{mathtools}
\usepackage{epsfig}
\usepackage{color}
\usepackage{url}
\usepackage{siunitx}
\usepackage{enumitem}

\usepackage{mathptmx}

\begin{document}

\preprint{AIP/123-QED}

\title[Multi-state MRAM cells for hardware neuromorphic computing]{Multi-state MRAM cells for hardware neuromorphic computing}
% Force line breaks with \\

\author{Piotr Rzeszut}
\email{piotrva@agh.edu.pl}

\author{Jakub Ch\ifmmode \mbox{\k{e}}\else \k{e}\fi{}ciński}
\author{Ireneusz Brzozowski}
\author{Sławomir Zi\ifmmode \mbox{\k{e}}\else \k{e}\fi{}tek}
\author{Witold Skowroński}
\affiliation{AGH University of Science and Technology, Institute of Electronics,\\ Al. Mickiewicza 30, 30-059 Krak\'{o}w, Poland}%
%\author{Jerzy Wrona}
%\affiliation{Singulus technologies, Kahl am Main, 63796, Germany}%
\author{Tomasz Stobiecki}
\affiliation{AGH University of Science and Technology, Institute of Electronics,\\ Al. Mickiewicza 30, 30-059 Krak\'{o}w, Poland}%
\affiliation{AGH University of Science and Technology, Faculty of Physics and Applied Computer Science,\\Al. Mickiewicza 30, 30-059 Krak\'{o}w, Poland}%

\date{\today}% It is always \today, today,

\begin{abstract}
Magnetic tunnel junctions (MTJ) have been successfully applied in various sensing application and digital information storage technologies. Currently, a number of new potential applications of MTJs are being actively studied, including high-frequency electronics, energy harvesting or random number generators. Recently, MTJs have been also proposed in designs of a new platforms for unconventional or bio-inspired computing. In the present work, it is shown that serially connected MTJs forming a multi-state memory cell can be used in a hardware implementation of a neural computing device. A behavioral model of the multi-cell is proposed based on the experimentally determined MTJ parameters. The main purpose of the mutli-cell is the formation of the quantized weights of the network, which can be programmed using the proposed electronic circuit. Mutli-cells are connected to CMOS-based summing amplifier and sigmoid function generator, forming an artificial neuron. The operation of the designed network is tested using a recognition of the hand-written digits in 20 $\times$ 20 pixel matrix and shows detection ratio comparable to the software algorithm, using the weight stored in a multi-cell consisting of  four MTJs or more.
\end{abstract}

\maketitle

\section{\label{sec:Introduction}Introduction}
Unconventional computing architectures such as artificial neural networks (ANN) have superior properties over conventional CMOS-based circuits in solving a number of computational problems, e.g., image or voice recognition, navigation, optimization and prediction\cite{fu2017look, venayagamoorthy1998voice, zhang2017type, muralitharan2018neural, abhishek2012weather}. As a concept, neural networks have been proved to be fast, flexible and energy-efficient. However, their digital implementation uses large amount of resources\cite{nurvitadhi2016accelerating}, which leads to high area needed to implement them. 
An alternative solution, opposite to the digital implementation, is to use analog-based circuits, where signals are represented as continuous voltage values rather than quantized bits\cite{yao2020fully,yao2017face,yu2018neuro,ambrogio2018equivalent}. In such implementations, a key element is a programmable resistive element, such as memristor\cite{strukov2008missing}, which can act as a weight in an artificial neuron. Using a solely digital implementation of a neural network may lead to high resource and energy consumption. On the contrary, using mixed digital and analog electronic circuits may enable more compact and energy efficient solution.  
In a number of the proposed analog ANN implementations, neuron behavior was mimicked by a resistive RAM (RRAM) element \cite{burr2017}, whose resistance change originated from the conductor/insulator transition\cite{yao2020fully}. However, cells based on resistive or phase-change technology suffer from limited durability and may degrade over time and subsequent programming cycles\cite{wu2018improvement}. On the contrary, spintronic elements such as memristors, nano-oscillators \cite{grollier2016} or probabilistic bit \cite{borders2019}, based on magnetic tunnel junctions (MTJ), which rely on the magnetisation switching or dynamics, do not have such endurance issues, are compatible with the CMOS technology and have been already shown to exhibit superior biomimetic properties \cite{romera2018}. In addition, recent theoretical works have predicted that neural networks are able to work efficiently not only with weights represented by real numbers but also with binary or quantized values\cite{moons2017minimum,hubara2017quantized,toledo2019mtj}.

Recently, we have proposed a design of multi-state spin transfer torque magnetic random access memory (STT-MRAM) cells\cite{rzeszut2019multi}, which may be used in neuromorphic computing schemes as synapses\cite{zhang2016all, torrejon2017neuromorphic, lequeux2016magnetic, sung2018perspective, sulymenko2018ultra, fukami2018perspective} or as a standard multi-state memory unit. In this paper, we present a fully functional hardware implementation design of a neural network, which needs no additional components for operation, except for input and output devices. The design of a single synapse is based on multi-bit STT-MRAM cells, interconnected with a minimal set of transistors forming amplifiers in conventional CMOS technology. The entire network is made of neurons arranged in four layers. The operation principle of the proposed neural network is validated using handwritten digits recognition task utilizing MNIST\cite{deng2012mnist} database. 

\section{\label{sec:experimental}Experimental and circuit level simulations}
%\subsection{\label{sec:multibitSimulation}Multi-bit MTJ cell - experiment and simulation}
In order to perform simulations of the proposed neural network prior to the fabrication process, critical parameters of serially connected MTJs forming a programmable memristor device are needed. 
In theory, such parameters could ultimately be derived from a fully-developed simulation of an MTJ dynamics, based on the Landau-Lifshitz-Gilbert equation. However, this would require an introduction of multiple assumptions on physical parameters such as magnetic anisotropy, spin-current polarization or temperature dependence, making the obtained result less general. On the other hand, the only thing that an MRAM cell needs in order to operate as a neural network weight is it's transfer function - or, equivalently, the resistance vs. voltage ($R(V)$) loop. Since our aim is to model the behavior of a realistic ANN consisting of thousands of individual MTJs, we find it less computationally expensive and more universal to model $R(V)$ loops directly, following the approach described below. We will show that this approach is corroborated by experimental data and that it provides a robust foundation for building a functional ANN.

\subsubsection{\label{sec:SingleModel}Modeling of single MTJ $R(V)$ curve}
\begin{figure}[h!]
\begin{center}
	\includegraphics[width=\columnwidth]{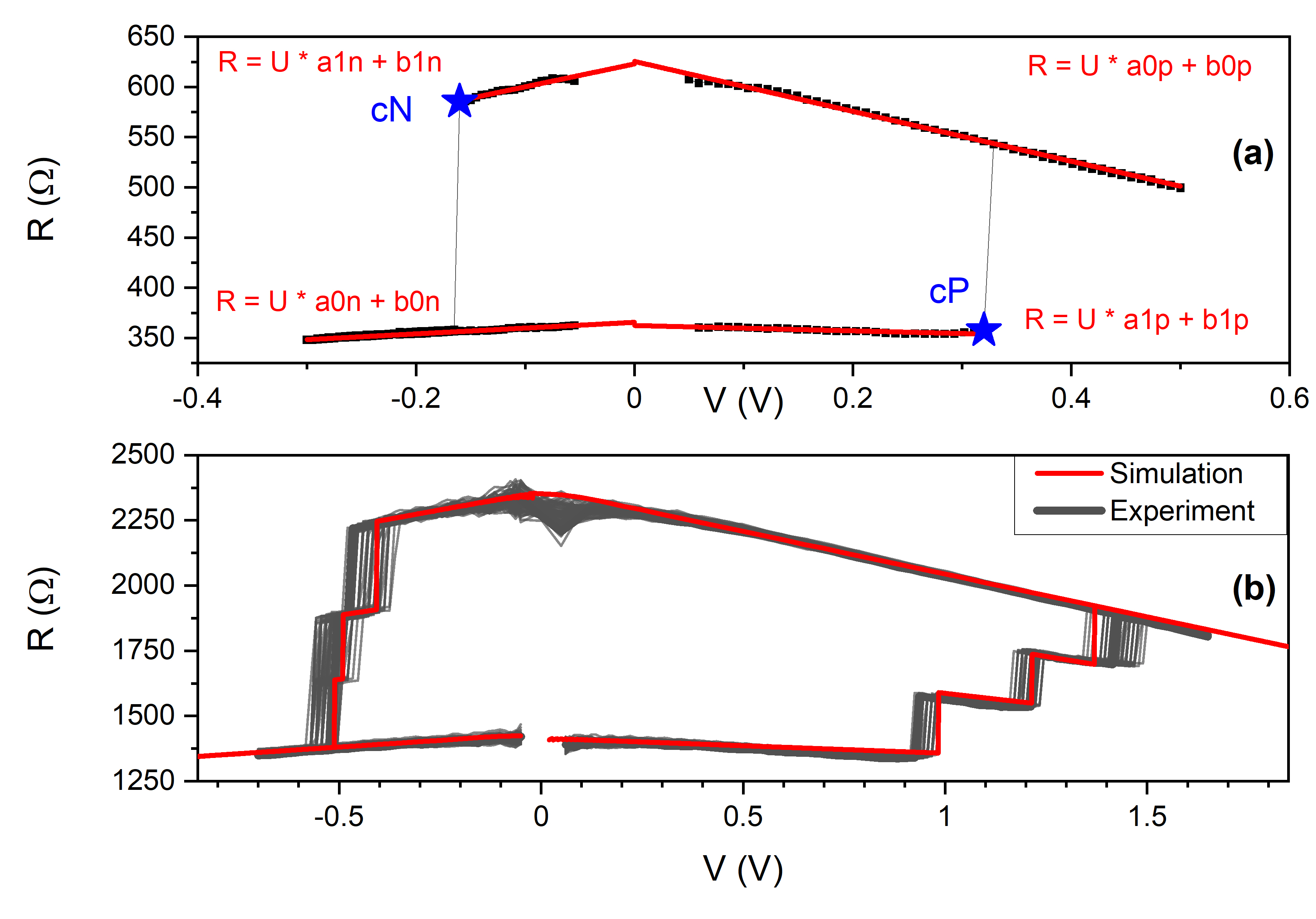}
	\caption{(a) Experimental $R(V)$ dependence (solid points) and the model consisting of four lines and two critical points (stars) presenting a single MTJ behaviour. (b) A representative simulation result of three serially-connected MTJs (solid line) together with a series of exemplary measurements (gray-scale lines) of a two-bit mutlicell. Parameters of a single MTJ were used to model the multi-cell characteristics.}
\label{fig:SingleModel}
	\end{center}
\end{figure}

A typical $R(V)$ loop of an MTJ may be approximated using four linear functions (resistance vs. bias voltage dependence in each MTJ state) and two threshold points (switching voltages) as presented in Fig. \ref{fig:SingleModel}. In addition, in the case of a real MTJ the following parameters are related to each other: $a1n = -a1p = a1$, $b1n = b1p = b1$, $a0n = -a0p = a0$ and $b0n = b0p = b0$. Moreover, a current resistance state (high or low resistance) has to be included.
Using such a model of the $R(V)$ curve allows also to calculate other transport curves, including $V(I)$. The proposed model corresponds to all MTJs that were investigated during the study.

%\subsubsection{\label{sec:Simulation}Simulation of serially connected MTJs}
Due to the bias dependence of a tunneling magnetoresistance effect \cite{sun_2008}, the resistance of an MTJ varies with voltage (thus $V(I)$ is not linear) and there is no direct way of calculating $R(V)$ transport curve for serially connected MTJs. Instead, $V(I)$ may be calculated using Kirchhoff's law, due to the fact that in serially connected MTJs the current is constant, while total voltage is a sum of the voltage drops on each element \cite{fiorentini2020}. In order to determine a resistance for a given voltage $V$, the following steps are performed\cite{SerspinSimRepo}:

\begin{enumerate}[itemsep=0pt,parsep=0pt]
	\item Start with a given state of all MTJs and current $I = 0$
	\item Increase current by a small step $I = I + \Delta I$ \label{it:start}
	\item For each MTJ:
	\begin{enumerate}[itemsep=0pt,parsep=0pt,leftmargin=\parindent, topsep=0pt, partopsep=0pt]
	\item Check if critical current for the given state was reached:
	\item If \textit{yes}, set $I$ to 0, change state of the element and go to \ref{it:start}. If \textit{no}, continue.
	\item Calculate voltage $V_i$ on the MTJ for given $I$
	\end{enumerate}
	\item Calculate total voltage $V_t = \sum_{i=1}^{N}V_i$
	\item If total voltage $V_t \geqslant V$, the solution was found and the algorithm ends. Otherwise, return to step \ref{it:start}.
\end{enumerate}

Depending on the polarity of the applied voltage, the sign of $\Delta I$ may be the same or opposite to the sign of applied $V$. The value of $\Delta I$ should be as small as possible for a given computation time limit. By repeating the process for each requested $V$, a full $I(V)$ and $R(V)$ curve can be calculated. 

Our approach makes it very straightforward to account for the fact that real-life MTJs are never fully identical, but instead have slightly different dynamical behavior due to fabrication imperfections and partially stochastic nature of the current induced magnetisation switching process \cite{Fukushima_2014}. Thanks to the fact that the operational principle of each multi-MTJ cell is described by it's $R(V)$ curve, it is sufficient to represent this variation by choosing slightly different $R(V)$ parameters for each cell at the beginning of the simulation. 

To verify the proposed model, a series of measurements of $R(V)$ curve on three different elements were repeated several tens of times for each element, and the data was used to obtain parameters of the model. Then, a series of measurements of these MTJs connected in series were performed. The comparison of experimental data and simulation results of the exact same MTJs is presented in Fig. \ref{fig:SingleModel} (b). One can see that the simulation results follow the experimental dependence closely, with even multistep-switching from high- to low-resistance state being reproduced.

%\subsubsection{\label{sec:SpreadAnalysis}Analysis of manufacturing spread}
In order to predict behavior of a mutli-cell, measurements of $R(V)$ loops for a few tens of fabricated MTJs were conducted and analysed. Next, for each measurement (Fig. \ref{fig:SingleModel}(a), the model was fitted, resulting in estimation of fabrication parameters and process yield. Across the sample these parameters represented a normal distribution and, therefore, the expected values $\mu$ and their standard deviation $\sigma$ may be used to model each parameter. 

\begin{table}
\caption{\label{tab:paramsSpread}Expected values and standard deviation of parameters used for simulations}
%\begin{ruledtabular}
\begin{tabular}{ccccc}
Param.&Unit&\mbox{$\mu$}&\mbox{$\sigma$}\\
\hline
a1&\si{\per\ampere}&-310&3\\
b1&\si{\ohm}&665&12\\
a0&\si{\per\ampere}&-30&3\\
b0&\si{\ohm}&360&12\\
cN&\si{\ampere}&-3.1e-4&1.5e-5\\
cP&\si{\ampere}&8.0e-4&1.5e-5
%\footnote{Negative critical current}
\end{tabular}
%\end{ruledtabular}
\end{table}

\begin{figure}[h!]
\begin{center}
	\includegraphics[width=\columnwidth]{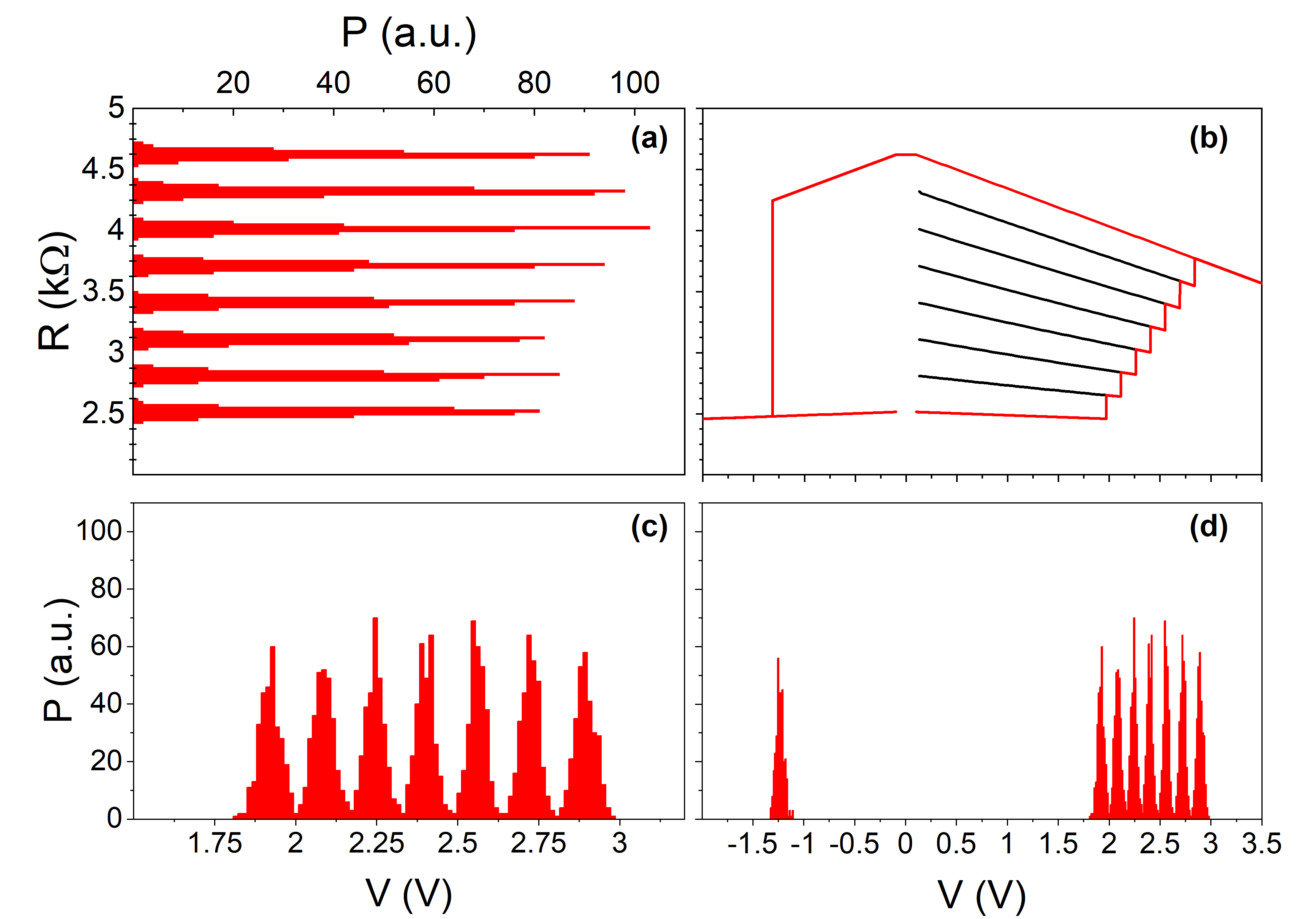}
	\caption{Simulation results for seven serially connected MTJs with given parameter spread. (a) Spread of readout resistances for the simulation. (b) Representative write-read-erase curves. Red line represents full write-read-erease cycle, while black ones represent write-read cycles while programming subsequent values. (c) Spread of positive write voltages for the simulation. (d) Spread of all write voltages for the simulation.}
\label{fig:SpreadSim}
	\end{center}
\end{figure}

The parameters presented in Tab. \ref{tab:paramsSpread} were used to simulate a mutli-cell composed of seven serially connected MTJs, where eight stable resistance states may be observed. After repeating the simulation 300 times, distribution of the switching (write) voltages and resistances were calculated (Fig. \ref{fig:SpreadSim}). A clear separation between stable resistance levels (readout values) as well as between write voltages were observed.

\subsection{\label{sec:synapse}Electronic neuron}
\begin{figure}
\begin{center}
	\includegraphics[width=\columnwidth]{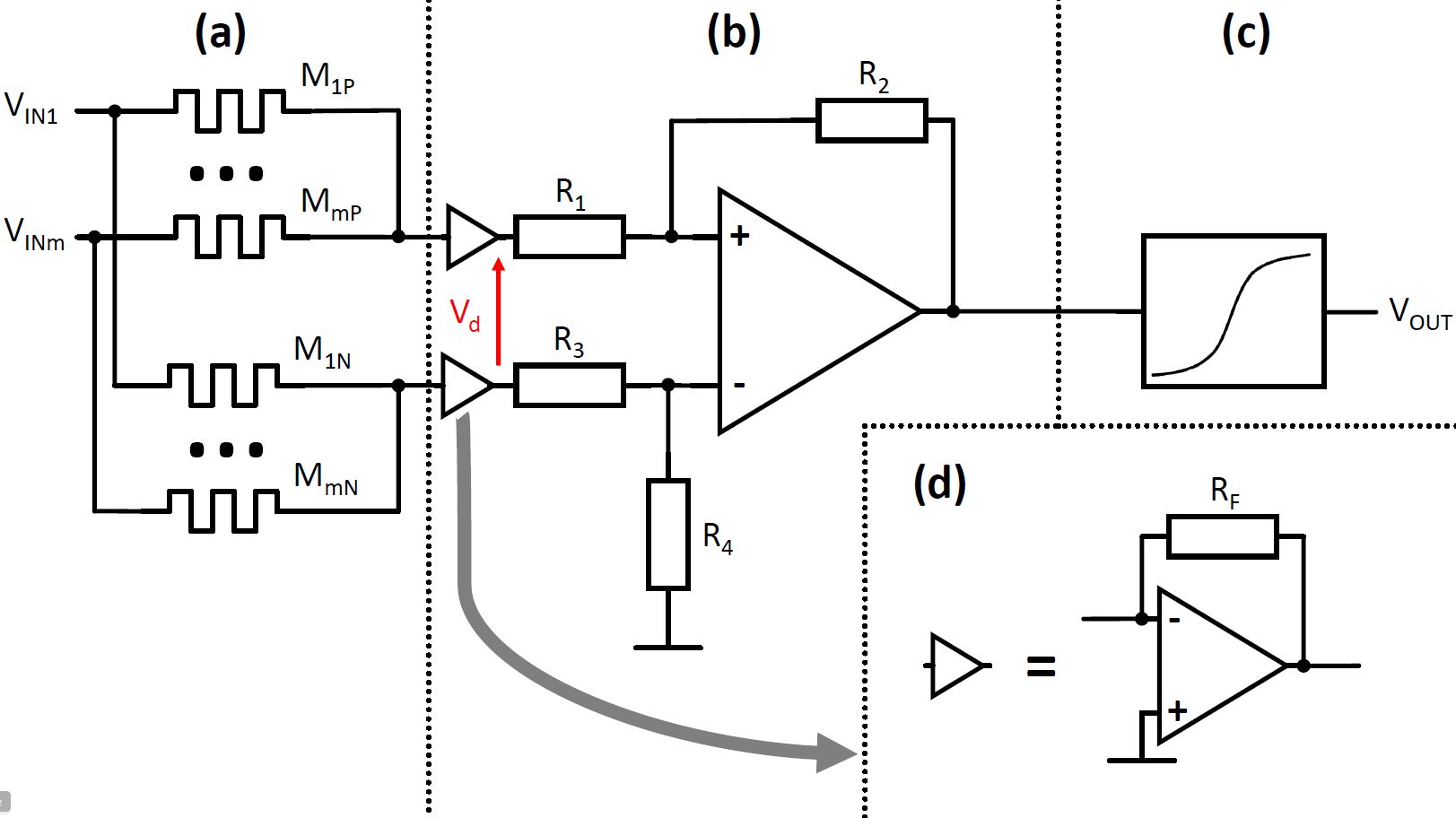}
	\caption{The proposed neuron design with multi-cells. The circuit consist of (a) a set of memristors for quantized weight purpose, (b) differential amplifier with summing amplifiers (d) at input and (c) sigmoid function block.}
\label{fig:WeightedSumIdea}
	\end{center}
\end{figure}

After the analysis of the multi-cell, which may be used as programmable resistor for performing weighted sum operation for many input voltages, we turn to the artificial neuron design. A schematic diagram of the proposed neuron is presented in Fig. \ref{fig:WeightedSumIdea}. Inputs and output ($V_{INm}$, $V_{OUT}$) are provided as bipolar analog signals. To enable positive and negative weights, each of the signal inputs uses a pair of programmable MTJ multi-cells ($M_{mP}$ and $M_{mN}$). In the case when the multi-cell resistances meet the condition $M_{mP} < M_{mN}$, a positive weight is achieved, whereas for the case of $M_{mP} > M_{mN}$ a negative weight value is realised. An alternative design with multiple MTJs connected in series with a separate select transistors has been proposed recently in Ref. \onlinecite{amirany2020}. For an equal mutli-cells' resistances, a zero weight is provided, which is equivalent to the situation when an input is disconnected from the synapse. The summing amplifier architecture is being used in order to realise addition operation while reducing footprint of the synapse. A differential amplifier converts differential voltage to a single bipolar signal, which is transformed using a non-linear (sigmoid) function. This voltage may be used as the input of the next synapse, or as the output of the network. Additionally, to provide a constant bias, a standard input with constant voltage may be used, where the level of this constant bias is determined in the same way as weights for other functional inputs.

\subsection{\label{sec:circuit}Neural network circuit}
The electrical circuit realizing proposed neural network was designed in a standard CMOS technology – UMC \SI{180}{\nano\metre}. To program the demanded resistance of seven serially connected MTJs, a voltage of about \SI{3.25}{\volt} is needed, so input/output (I/O) \SI{3.3}{\volt} transistors were used to design a circuit for MTJs programming purpose, while for other circuits, a standard \SI{1.8}{\volt} transistors were used. An individual neuron circuit is composed of three parts. At the input, two resistive networks consisting of memristors realise a multiplication of input voltages and coefficients - Fig. \ref{fig:WeightedSumIdea}(a). Next, the obtained voltage is amplified to the demanded value in a differential amplifier - Fig. \ref{fig:WeightedSumIdea}(b). Finally, the third part is a sigmoid function block, which realises the activation function - Fig. \ref{fig:WeightedSumIdea}(c). The circuit for memristors programming is designed using I/O transistors (not shown) and consists of high voltage switches and digital elements to control them. 

The differential voltage $V_d$ generated by the divider network (Fig. \ref{fig:WeightedSumIdea}(a)) connected to a pair of summing amplifiers (Fig. \ref{fig:WeightedSumIdea}(d)) can be expressed as:
$$
V_d = -R_f(\sum_{i=1}^{m}V_{INi}(G_{iP}-G_{iN})),
$$
where:
$$
G_{iX} = \frac{1}{M_{iX}}
$$

\begin{figure}
\begin{center}
	\includegraphics[width=\columnwidth]{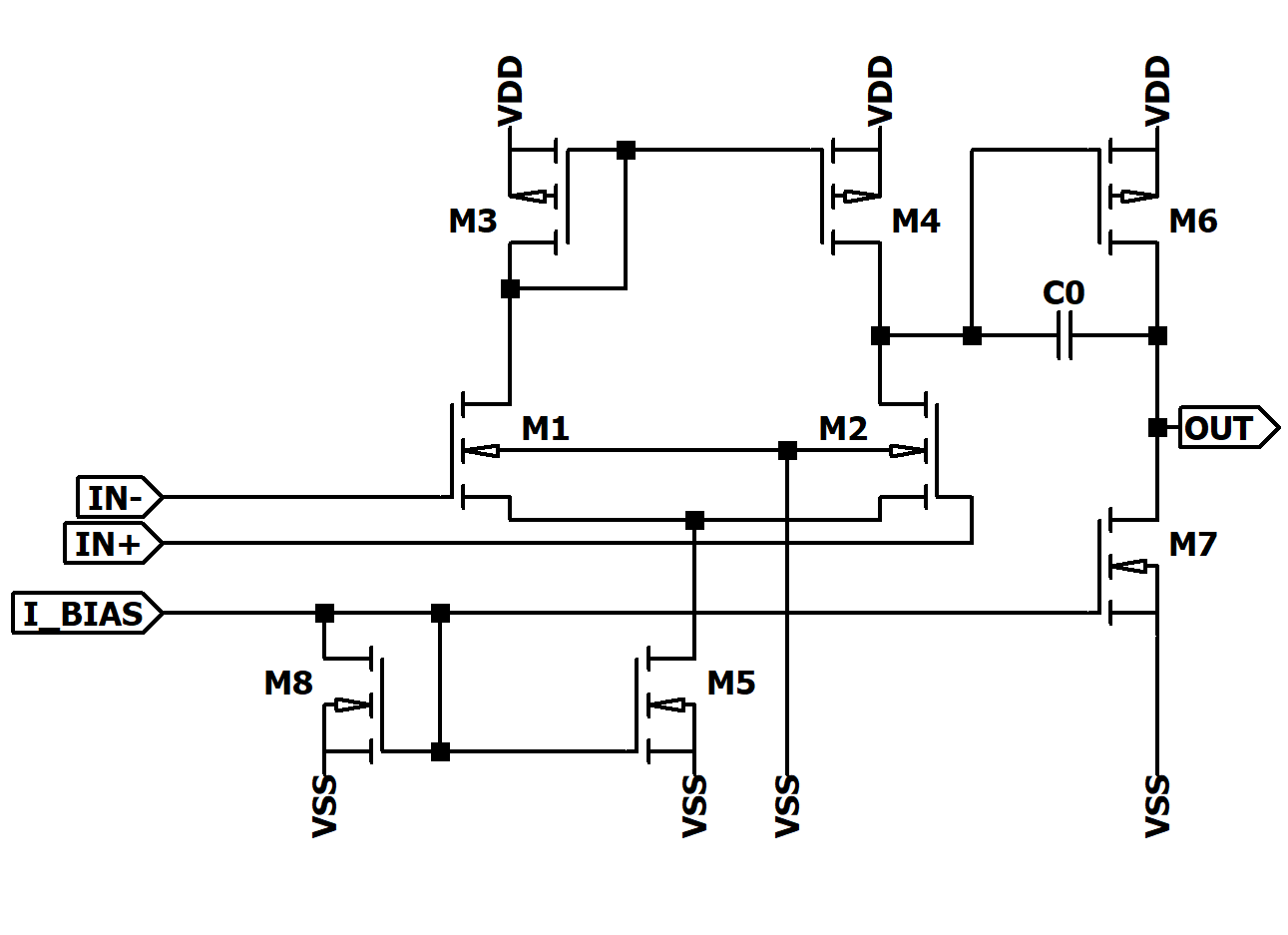}
	\caption{A proposed differential amplifier circuit.}
\label{fig:amplifier}
	\end{center}
\end{figure}

The operational amplifier, presented in Fig. \ref{fig:amplifier}  was designed as two stage circuit consisting of a differential pair M1, M2 with a current mirror load M3, M4 biased by M5 with a current of \SI{1}{\micro\ampere}. The output stage M6, M7 provide appropriate amplification and output current. The total current consumed by the operational amplifier is about \SI{12}{\micro\ampere} and amplification with an open loop of around \SI{74}{\dB}. Dimensions of transistors were chosen in such a way to obtain the smallest area possible while meeting the required electrical parameters. 

\begin{figure}
\begin{center}
	\includegraphics[width=\columnwidth]{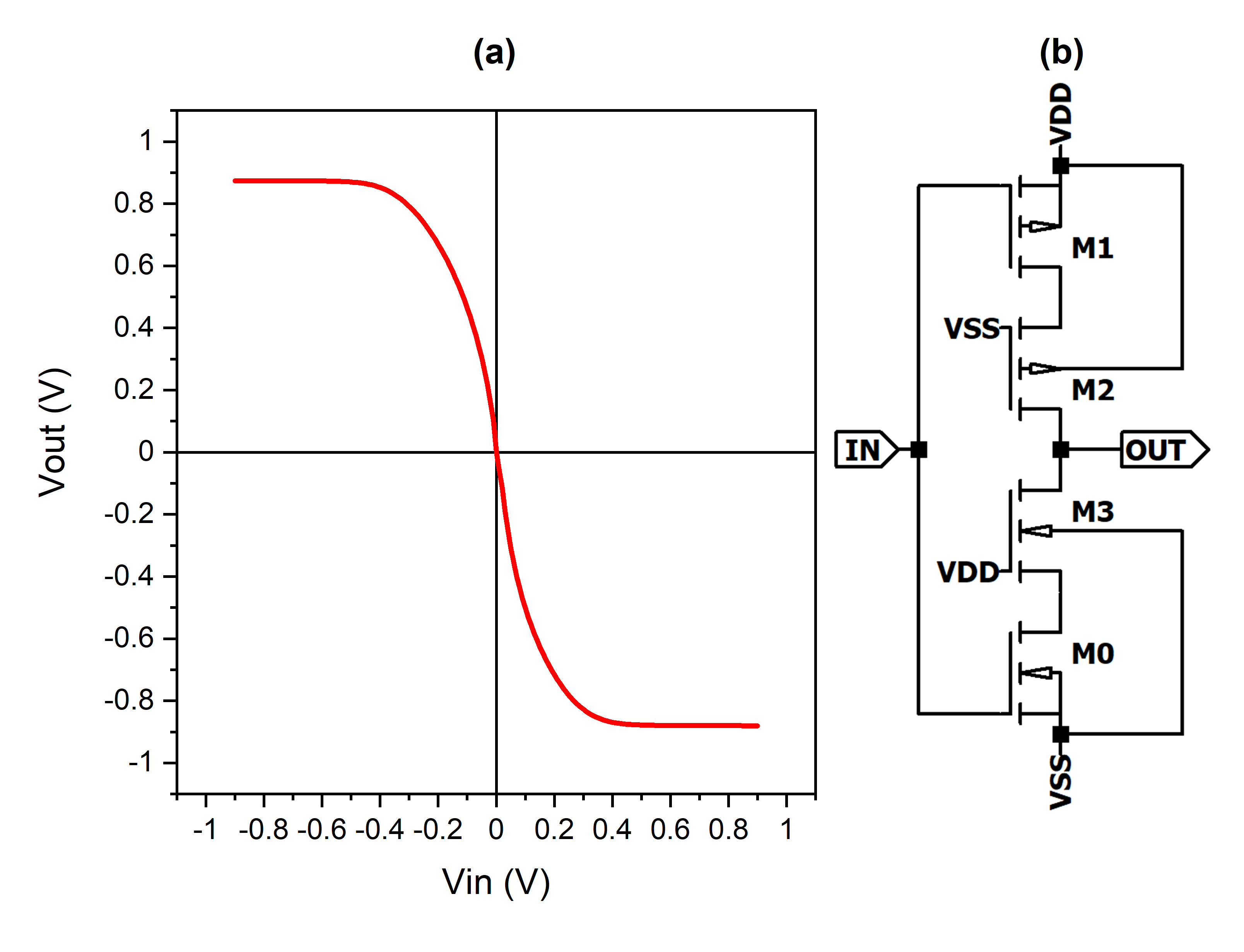}
	\caption{(a) A transfer function of a sigmoid-generating inverter realized by (b) the proposed inverter circuit.}
\label{fig:inverter}
	\end{center}
\end{figure}

The final stage of the neuron is a circuit, which performs activation functions and has sigmoidal transfer characteristic, presented in Fig. \ref{fig:inverter}(b). It is designed as a modified inverter \cite{Khodabandehloo2011}. Transistors M2 and M3 work as resistors, moving operating point of transistors M0 and M1 to the linear region. Finally, the circuit realizes the transfer characteristic shown in Fig. \ref{fig:inverter}(a). Minimum length of channels were used (\SI{180}{\nano\metre}, except for M3, which uses \SI{750}{\nano\metre}), while their width was chosen to obtain required characteristics and output current necessary to drive the next stage. Therefore, the width of M0 and M3 is \SI{1.2}{\micro\metre}, M1 is \SI{4.56}{\micro\metre}, and M2 is \SI{1.12}{\micro\metre}. 

\section{\label{sec:results}Results and discussion}

\begin{figure}
\begin{center}
	\includegraphics[width=\columnwidth]{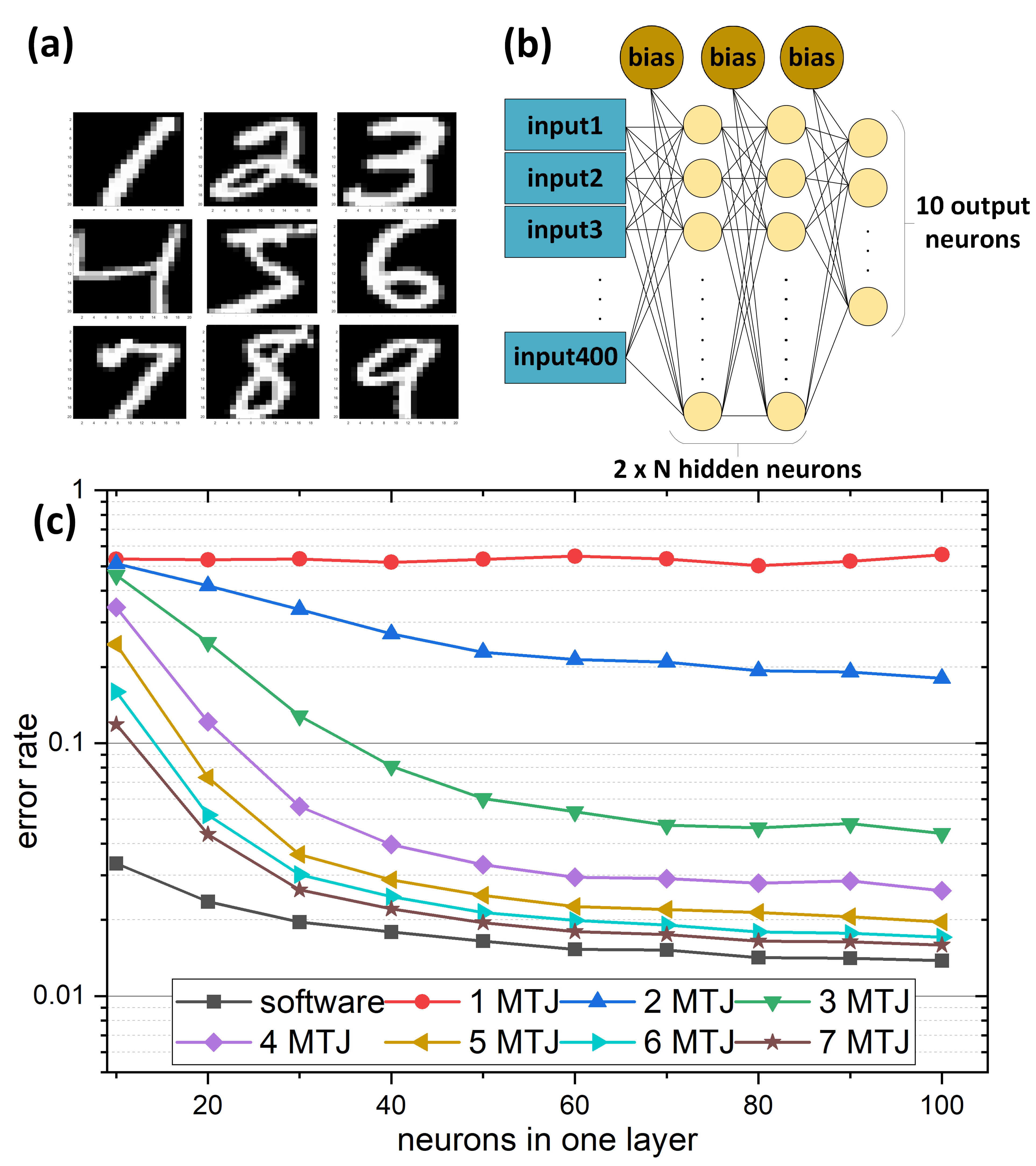}
	\caption{Simulated neural network based on multi-bit MRAM cells. Handwritten digits from MNIST database (a) are recognized by a standard neural network with architecture shown in (b), where black lines represent network weights and yellow circles represent individual neurons. After training, weights calculated by software are replaced by discretized values corresponding to 1-7 serial MTJs MRAM cells, which affects the network performance (c).}
\label{fig:NeuralNetwork}
	\end{center}
\end{figure}
To evaluate the performance of the multi-bit MTJ cell-based ANN a set of classification tasks using the MNIST dataset of handwritten digits (Fig.\ref{fig:NeuralNetwork}(a)) was prepared. The conceptual architecture used for the network is shown in Fig. \ref{fig:NeuralNetwork}(b) and consists of the input layer, two hidden layers containing $N$ neurons each and the output layer. The network was trained using the standard scaled conjugate gradient method and cross-entropy error metrics, with \textit{tanh} activation function for every layer except the last one, where the \textit{softmax} function was used. Then, its performance was evaluated on a testing subset that has been drawn randomly from the input data and has not participated in training. This procedure was repeated 50 times in total, with training and testing subsets being reshuffled each time, leading to an average error estimate for each network size.
\\
Having established the benchmark network, the evaluation of our MTJ-based design was performed. The original float-accuracy weights between different neurons were replaced by a discrete version corresponding to multi-state MTJ synapses and characteristics of designed amplifier and activation function circuit were applied to the model. The new weights were calculated using simulated conductance data (as described in Sec. \ref{sec:experimental}) and rescaled by tuning amplifiers' gains to match the desired value range for the neurons. Then, the performance of the network was re-evaluated on the testing data subset. The results are presented in Fig. \ref{fig:NeuralNetwork}(c). It can be seen that, as long as the number of MTJs used per multi-state cell exceeds three, the performance of the MTJ-based solution is comparable to the original software version, with differences being only incremental in character.

\section{\label{sec:summary_final}Summary}
The presented architecture of full hardware artificial neural network proves to be an effective way of performing neuromorphic computing. Compared to other solutions, it utilizes standard MTJs that are compatible with STT-MRAM technology, which has been recently developed for mass production. Additionally, MTJs in such application are very stable over time and they exhibit high endurance in terms of reprogramming, comparing to low-energy barrier MTJs used in probabilistic computing. Moreoever, the presented solution enables more efficient computing, as devices may benefit from a multi-state memristor. To validate this, the artificial CMOS-based neuron was designed, consisting of multi-cell based synapses, differential amplifiers and sigmoid function generator. It was shown that the quantized-weight approach enables the developement of a functional artificial neural network, capable of solving recognition problems with accuracy level similar to the benchmark software model.

\section*{Acknowledgement}
We would like to thank PhD J. Wrona from Singulus Technologies AG for MTJ multilayer deposition. Scientific work funded from budgetary funds for science in 2017-2018, as a research project under the "Diamond Grant" program (Grant No. 0048/DIA/2017/46).  W.S. acknowledges support by the Polish National Center for Research and Development grant No.LIDER/467/L-6/14/NCBR/2015. T.S. acknowledges the SPINORBITRONICS grant No. 2016/23/B/ST3/01430. The nano-fabrication process was performed at Academic Centre for Materials and Nanotechnology (ACMiN) of AGH University of Science and Technology.

%\nocite{*}
\bibliography{bibliography}% Produces the bibliography via BibTeX.

\end{document}